\documentclass[a4paper]{article}

\usepackage{amsmath}
\usepackage[dvips]{graphicx}
\usepackage[latin1]{inputenc}
 \def \d {\mathrm{d}}
\def \ni {\noindent}

\def\scri{
\unitlength=1.00mm
\thinlines
\begin{picture}(3.5,2.5)(3,3.8)
\put(4.9,5.12){\makebox(0,0)[cc]{$\cal J$}}
\end{picture}}
\begin{document}

\begin{center}
{\Large Conformal isometry of the Reissner-Nordström-de Sitter black hole} \\
\vspace{4ex}
 
{\large Johan Brännlund}\footnote{E-mail address: jbr@physto.se}
 
\vspace{2ex}
 
Fysikum\\
Stockholm University\\
AlbaNova \\
S-106 91 Stockholm \\
Sweden
 
\vspace{2ex}

  {\bf Abstract} \\
\end{center}

\ni It was pointed out by Couch and Torrence that the extreme
Reissner-Nordström solution possesses a discrete conformal isometry.
Using results of Romans, it is shown that such a symmetry also exists
when a non-zero cosmological constant is allowed.

\section*{Introduction}
In \cite{ct}, Couch and Torrence found a conformal isometry that
interchanges the event horizon and null infinity $\scri$ of an
extremal Reissner-Nordström black hole. Related but distinct ideas
have also appeared in a string theory context\cite{gary}. Here we will
show that such a conformal isometry exists also for the case of a
positive cosmological constant, provided that the surface gravities of
the two horizons are equal.  We will leave global issues aside and
refer the reader to \cite{bh} for those matters. See also \cite{gt}
for a discussion of conserved quantities in asymptotically de Sitter
spacetimes from a Hamiltonian point of view.

\section*{The calculation}
In the Couch-Torrence case the conformal isometry switches the roles
of the event horizon and infinity. With a positive cosmological
constant, there is another geometrically distinguished object: the
cosmological horizon. As $\Lambda \rightarrow 0$, this horizon
approaches infinity.  Let us therefore provisionally assume that the
putative conformal isometry interchanges the black hole horizon and
the cosmological horizon since this produces the correct limiting
behavior; this assumption will later turn out to be correct.

Assume that the BH horizon is at $r=a$ and the cosmological horizon at
$r=b$. It is convenient to introduce the coordinate $x$ defined by

\begin{displaymath}
  x = \frac{r-a}{b-r} \frac{b}{a}
\end{displaymath}

The strategy here is to first re-express the above metric in terms of
the $x$ coordinate.  In such a coordinate system, the metric will be
manifestly conformally invariant under the inversion $x \mapsto 1/x$
as we will now show (note that the surface gravities of the two
horizons will remain equal after the inversion since the surface gravity
$\kappa$, suitably defined, is a conformal invariant\cite{jk}).

For a general spherically symmetric black hole, the metric
in the standard coordinates is

\begin{equation}
\label{bhmetric}
  \d s^2 = -V(r) \d t^2 + \frac{\d r^2}{V(r)} + r^2 \d \Omega^2
\end{equation}

\ni where $V(r)$ is given by

\begin{equation}
\label{vfunc}
  V(r) = \frac{1-2m}{r}+\frac{Z^2}{r^2}-\frac{1}{3}\Lambda r^2
\end{equation}

A calculation shows that $\kappa=\vert V' / 2 \vert$. For the
particular case of so-called ``lukewarm black holes'', where the
surface gravities of the two horizons are equal, Romans has
shown\cite{romans} that the parameters take the values
\begin{displaymath}
  m=\frac{ab}{a+b}=\pm Z \hspace{3ex} \Lambda=\frac{3}{(a+b)^2}
\end{displaymath}

Note that the mass remains equal to the charge when a cosmological
constant is introduced, so in this sense the black hole is still
extremal. Inserting those parameter values into $V(r)$, one obtains

\begin{displaymath}
  V(r) \propto \frac{1}{r^2} (r-a)(r-b)(r^2+(a+b)r-ab)
\end{displaymath}
In terms of $x$, the right-hand side of (\ref{vfunc}) becomes
\begin{equation}
\label{dub}
  W(x)=\frac{x}{(1+x)^2(b+ax)^2} \left( 2ab(1+x^2)+x(a+b)^2 \right)
\end{equation}

We are now in a position to verify that the map $x \mapsto 1/x$ is
indeed a conformal isometry. The condition for this to hold is that
$\d s^2(x) = f^2 \d s^2(1/x)$ where $f^2$ is the conformal factor.
The easiest way to verify this is to notice that the $\d \Omega^2$
term in the metric only involves $x$ through the $r^2$ so we
immediately find that the conformal factor $f^2$ has to satisfy

\begin{equation}
\label{conf}
  f^2 = \left( \frac{a+bx}{b+ax} \right)^2
\end{equation}

One can check that the conformal isometry conditions for the $\d t^2$
and the $\d r^2$ terms both lead to
\begin{displaymath}
  \frac{W(x)}{W(1/x)}=f^2
\end{displaymath}

Since $f^2$ is known from (\ref{conf}) and $W(x)$ is given by
(\ref{dub}), one can simply calculate the left-hand side of the above
expression and find that the equality does hold, meaning that the
inversion is indeed a conformal isometry.

\section*{Optical geometry}
All this can be nicely illustrated in terms of optical geometry\footnote{For
a more complete exposition of optical geometry for Reissner-Nordström
black holes, see \cite{optgeom}. For optical geometry in general, see \cite{ab}
or \cite{gb}.}. The optical
geometry is obtained as the spatial part of the conformally rescaled metric

\begin{displaymath}
  \d \tilde{s}^2 = \frac{1}{V(r)}\d s^2=
- \d t^2 + \frac{\d r^2}{V(r)^2} + \frac{r^2}{V(r)} \d \Omega^2
\end{displaymath}

\ni A null geodesic in the full spacetime projects down to a geodesic in
the optical geometry, hence its name.

\begin{figure}[h]
  \begin{center}
    \includegraphics[height=6cm]{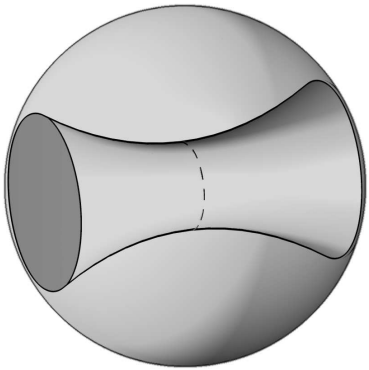}
     \caption{ Optical geometry of the R-N-dS black hole with one
       dimension suppressed, embedded in the Poincaré ball. The dashed
       line marks the location of the neck at $r=2m$. \label{optgeom} }
  \end{center}
\end{figure}

The optical geometry looks roughly like figure \ref{optgeom}. As
outlined above, we would expect the conformal inversion to interchange
the two horizons; in this picture, this would amount to swapping the
two spheres at infinity (they appear as circles in the figure). The
neck between the spheres is where the optical geometry has its minimal
radius; just like in the Schwarzschild case, this occurs at $r=2m$,
i.e. at $r=2ab/(a+b)$. One can verify that this is a fixed point of
the conformal inversion $x \mapsto 1/x$.

A calculation reveals the curvature scalar of the optical geometry 
associated with \ref{bhmetric} to be

\begin{displaymath}
  R_{opt} =\frac{1}{2r^2} \left( 4V - 4V^2 -3r^2 V'^2  + 4rVV' + 
4r^2 V V'' \right)
\end{displaymath}

This means that $R_{opt} \rightarrow -3V'^2/2 = -6 \kappa^2$ as $r
\rightarrow a$. If we restrict the definition of a Couch-Torrence
symmetry to be an isometry of the optical geometry, one obtains in
this way a necessary condition for such a symmetry to exist: since
$R_{opt} \rightarrow 0$ when one approaches infinity for an
asymptotically flat black hole, $\kappa$ must equal zero at the
horizon in that case.

\section*{Conclusions and open ends}
One may ask the question whether there are other black-hole spacetimes
for which a conformal isometry of the Couch-Torrence
type exists.  This seems to be tricky to answer; since the isometry is
discrete, infinitesimal methods using Killing vectors are not likely
to be very useful. One could hope for the existence of a sufficient
condition something along the lines of the curvature scalar condition
referred to above. This remains to be investigated.

\section*{Acknowledgements}
I am grateful to Ingemar Bengtsson for suggesting the problem and
for providing advice and constructive criticism.
I would also like to thank Jörgen Nätterlund for the figure.


\begin{thebibliography}{99}
\bibitem{ct} Couch, W.E.\& Torrence, R.J., Gen. Rel. Grav. {\bf 16},
  789 (1984)
\bibitem{gary} Gibbons, G., Horowitz, G. \& Townsend, P.K., Class.Quant.Grav.{\bf 12}
  297-318 (1995)
\bibitem{bh} Brill, D. \& Hayward, S., Phys.Rev. D {\bf 50} 4914-4919
  (1994)
\bibitem{gt} Gomberoff, A. \& Teitelboim, C., hep-th/0302204
\bibitem{jk} Jacobson T. \& Kang, G., Class. Quant. Grav. {\bf 10},
  L201-L206 (1993)
\bibitem{romans} Romans, L.J, Nucl. Phys. B {\bf 383}, 395-415 (1992)
\bibitem{optgeom} Abramowicz, M., Bengtsson, I. , Karas, V. \& Rosquist, K.
  Class. Quant.  Grav. {\bf 12}, 3963 (2002)
\bibitem{ab} Abramowicz, M., Carter, B. \& Lasota, J., Gen. Rel. Grav. {\bf 20}
  1173 (1988)
\bibitem{gb} Gibbons, G., Nucl. Phys. B {\bf 472} 683 (1996)
\end{thebibliography}
\end{document}